\PassOptionsToPackage{unicode}{hyperref}
\PassOptionsToPackage{hyphens}{url}
\PassOptionsToPackage{numbers,sort&compress}{natbib}
\documentclass[
]{article}
\usepackage{xcolor}
\usepackage{amsmath,amssymb}
\usepackage{iftex}
\ifPDFTeX
  \usepackage[T1]{fontenc}
  \usepackage[utf8]{inputenc}
  \usepackage{textcomp} 
\else 
  \usepackage{unicode-math} 
  \defaultfontfeatures{Scale=MatchLowercase}
  \defaultfontfeatures[\rmfamily]{Ligatures=TeX,Scale=1}
\fi
\usepackage{lmodern}
\ifPDFTeX\else
\fi
\IfFileExists{upquote.sty}{\usepackage{upquote}}{}
\IfFileExists{microtype.sty}{
  \usepackage[]{microtype}
  \UseMicrotypeSet[protrusion]{basicmath} 
}{}
\makeatletter
\@ifundefined{KOMAClassName}{
  \IfFileExists{parskip.sty}{%
    \usepackage{parskip}
  }{
    \setlength{\parindent}{0pt}
    \setlength{\parskip}{6pt plus 2pt minus 1pt}}
}{
  \KOMAoptions{parskip=half}}
\makeatother
\usepackage{longtable,booktabs,array}
\usepackage{calc} 
\usepackage{etoolbox}
\makeatletter
\patchcmd\longtable{\par}{\if@noskipsec\mbox{}\fi\par}{}{}
\makeatother
\IfFileExists{footnotehyper.sty}{\usepackage{footnotehyper}}{\usepackage{footnote}}
\makesavenoteenv{longtable}
\usepackage{graphicx}
\makeatletter
\newsavebox\pandoc@box
\newcommand*\pandocbounded[1]{
  \sbox\pandoc@box{#1}%
  \Gscale@div\@tempa{\textheight}{\dimexpr\ht\pandoc@box+\dp\pandoc@box\relax}%
  \Gscale@div\@tempb{\linewidth}{\wd\pandoc@box}%
  \ifdim\@tempb\p@<\@tempa\p@\let\@tempa\@tempb\fi
  \ifdim\@tempa\p@<\p@\scalebox{\@tempa}{\usebox\pandoc@box}%
  \else\usebox{\pandoc@box}%
  \fi%
}
\def\fps@figure{htbp}
\makeatother

\makeatletter
 \let\@cite@ofmt\@firstofone
 \def\@biblabel#1{}
 \def\@cite#1#2{{#1\if@tempswa , #2\fi}}
\makeatother
\newlength{\cslhangindent}
\newlength{\csllabelwidth}
 {\begin{list}{}{%
  \setlength{\itemindent}{0pt}
  \setlength{\leftmargin}{0pt}
  \setlength{\parsep}{0pt}
  \ifodd #1
   \setlength{\leftmargin}{\cslhangindent}
   \setlength{\itemindent}{-1\cslhangindent}
  \fi
  \setlength{\itemsep}{#2\baselineskip}}}
 {\end{list}}
\usepackage{calc}

\usepackage{bookmark}
\IfFileExists{xurl.sty}{\usepackage{xurl}}{} 
\usepackage{float}
\usepackage{caption}
\usepackage[numbers]{natbib}
\usepackage{authblk}
\hypersetup{
  hidelinks,
  pdfcreator={LaTeX via pandoc}}

\title{\textbf{DeepAJM: Deep Association Joint Model for Irregularly Sampled data}}

\usepackage{array}   
\usepackage{authblk}

\author[1]{Barsha Halder}
\author[2]{Jeffrey A. Thompson}

\affil[1,2]{Department of Biostatistics \& Data Science, University of Kansas Medical Center, Kansas City, KS, USA}

\date{}

\begin{document}
\maketitle

{\centering \section*{\large Abstract}\par}

Joint Models simultaneously model longitudinal and survival outcomes, leveraging patterns in patients’ longitudinal trajectory to improve the prediction of survival outcomes. The classical parametric joint models, however, rely on fixed parametric assumptions, making them susceptible to bias under model misspecification, smaller sample sizes. We propose a deep joint model, DeepAJM that does not require any parametric assumptions, while retaining a partially interpretable, per-longitudinal-outcome association structure. The joint model uses an encoder-decoder (sequence-to-sequence) architecture to learn the latent structure in patients’ time-varying covariate trajectories. The model links the longitudinal processes to the survival processes through a learned interpretable association structure, in which each longitudinal output from the decoder gets remodulated by baseline covariates before it contributes to the risk scores from the survival head of the architecture. The model was evaluated on three datasets- a cardiovascular-disease EHR cohort, a primary biliary cirrhosis (PBC2) dataset, and a simulated dataset- against a classical parametric joint model, TransformerJM, DA-LSTM and a Cox-based survival-only model. All models were assessed using C-index, integrated brier score (IBS), time-dependent AUROC, and time-dependent AUPRC. Our model achieved the best discrimination in terms of the C-index, time-dependent AUROC, and AUPRC across all datasets.   

\section{Introduction}\label{introduction}

A Joint model analyzes longitudinal and survival data simultaneously,
rather than modeling them separately, in order to capture how
longitudinal information influences the risk of an event. The two
processes \(-\) longitudinal and survival are linked through a shared
latent effect, such that changes in a variable trajectory directly
inform, and are reflected in, the survival outcomes. In clinical
studies, joint models play a significant role in predicting survival
outcomes. Wulfsohn and Tsiatis (1997) first proposed jointly
modeling longitudinal and time-to-event data \cite{wulfsohn1997joint}. Later, Rizopoulos developed practical tools in R \cite{rizopoulos2012joint, jmbayes2} to make this computationally intensive framework accessible and
extended it to accommodate non-linear time-varying trajectories and
different functional forms for the association linking the longitudinal
and survival processes. However, the traditional parametric joint model
relies heavily on rigid parametric assumptions and conditional
independence assumption between the two processes given the shared
random effects \cite{rizopoulos2012joint, han2007parametric}. Furthermore, with higher dimensionality, the parametric approach suffers from computationally complexity as numerical integrations are required for the likelihood estimation.

Non-parametric deep-learning approaches have been proposed to overcome
these parametric constraints. Generally, these methods use a shared
temporal layer \(-\)recurrent (RNN/LSTM) or transformer-based that
exploits the temporal information in the longitudinal trajectory,
including its shape, and feeds both longitudinal and survival submodels.
To date, a few studies have explored joint models using deep
learning. TransformerJM \cite{TFJM}, a transformer-based architecture that jointly models longitudinal and survival data to predict survival probabilities. It discretizes time into a fixed set of intervals and
uses a binary discrete-time survival loss. However, this approach can be
dominated by multiple follow-up visits with no events, skewing
predictions toward higher survival and leading to poor discrimination. TransformerLSR, a continuous-time transformer model was proposed to jointly model longitudinal measurements, survival and recurrent events, and modeling survival and recurrent-event as flexible temporal point processes \cite{zhang2025transformerlsr}. However, rather than using an explicit association structure that link submodels as in classical joint models, TransformerLSR captures dependence implicitly through shared representations. DA-LSTM \cite{pmlr-v85-lim18a} also models longitudinal and time-to-event data using RNN-based architecture with a multi-task learning approach but relies on
discrete-step RNN and does not explicitly handle continuous, irregular
time gaps between consecutive observations. All these deep-learning
methods \cite{pmlr-v85-lim18a, TFJM, zhang2025transformerlsr, 10.1093/bioinformatics/btaa075, dynamicdeephit} jointly model longitudinal and survival data, but none link the longitudinal process to survival through an explicit shared effect as in a classical joint model \cite{rizopoulos2012joint}, and none models baseline covariates as outcome-specific modulation weights. Our study addresses both gaps and designs a joint model that processes longitudinal outcomes using a seq2seq architecture and survival processes with a semi-parametric cox formulation without discretizing event time into fixed intervals,
modulating each longitudinal outcome by baseline-covariates-specific
weights, and linking the two processes through an interpretable
association structure. Our proposed model, DeepAJM, comprises four
components: an LSTM-based seq2seq network with cross-attention
mechanism, a baseline-modulation network, an association structure
producing per outcome association weight, and a survival sub-network
that maps the association-weighted longitudinal outcomes to survival
risk scores.

\section{Methods}\label{methods}

The proposed joint model, DeepAJM, couples an attention-based
sequence-to-sequence longitudinal head with a baseline covariate
modulation network and a survival head. The longitudinal head, an
LSTM-based \cite{Hochreiter1997LongSM} encoder-decoder structure, processes the longitudinal
measurements and predicts the outcomes. These predicted outcomes are
then readjusted using output-specific weights learned by the modulation
network and linked to the survival head through an association
structure.

\subsection{Data preprocessing}\label{data-preprocessing}

\textbf{Input sequences}: The input sequences for the longitudinal head were constructed as four time points per subject, consisting of \(X\) longitudinal features, of shape \((N,\ T,\ X)\), where \(N\) is the number of subjects, \(T\) is the sequence length and \(X\) is the number of features. Both the encoder and decoder of the longitudinal head received the same input sequences of shape \((T,\ X)\) per subject. The longitudinal features consisted of clinical, physiological and functional biomarkers, along with the time at which each set of biomarkers was sampled. The time feature was encoded as time elapsed (in years) from the subject's baseline visit, allowing the model to account for irregular sampling intervals between visits. To ensure the longitudinal head received sufficient temporal information, each input sequence comprised four time points up to the landmark time \(\left( t_{L} \right)\). During training, \(t_{L}\) was subject's second-to-last visit \(\left( t_{n,\ T - 1} \right)\); during evaluation at fixed horizons \(\left( t_{pred} \right)\), \(t_{L}\) was the last visit preceding the earliest evaluation horizon, restricting the cohort to subjects with at least four visits up to \(t_{L}\) who remained at risk at \(t_{L}\).

\begin{figure}
    \centering
    \includegraphics[width=0.95\linewidth]{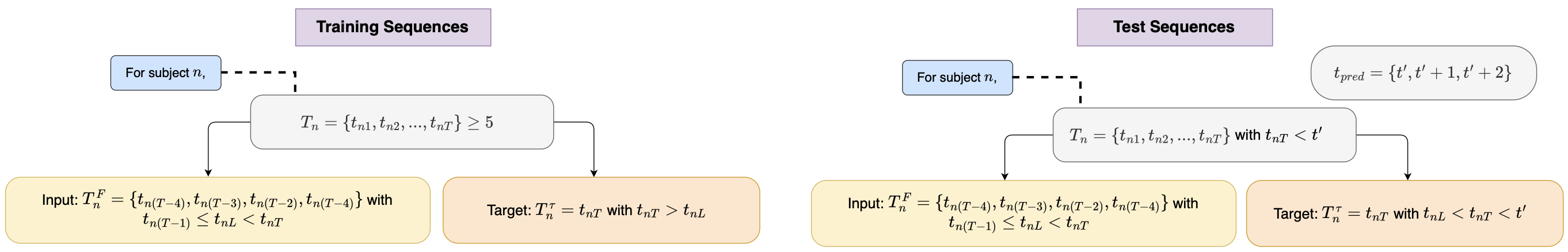}
    \caption{Construction of training and test sequences used across all evaluated methods.}
    \label{fig:seq-construction}
\end{figure}

\textbf{Baseline covariate representation}: The baseline covariates in our study
were age \(\left( b_{1} \right)\) and sex \(\left( b_{2} \right)\) and
were processed through baseline covariate modulation network. In
longitudinal studies, baseline covariates exert distinct effects on both
longitudinal and survival outcomes; although a deep joint model learns
those feature interactions internally across timesteps, it may not fully
capture the interaction between baseline covariates and longitudinal
outcomes. Age (\(b_{1})\), a continuous scalar was projected through a
dense layer with ReLU activation to explicitly capture its non-linear
effects on longitudinal trajectories, thereby improving survival
predictions; whereas sex \({(b}_{2})\) was retained as a binary
indicator. The projection dimension for age was selected via
hyperparameter tuning

\begin{equation}
{\widetilde{b}}_{1} = ReLU\left( W_{g}b_{1} + c_{g} \right)
\end{equation}

\subsection{Proposed Model}\label{proposed-model}

\subsubsection{\textbf{Longitudinal head}}

The longitudinal head was designed as an LSTM-based sequence-to-sequence
architecture augmented with a multi-head cross-attention mechanism, used
to extract dynamic temporal patterns from longitudinal trajectories.
LSTMs are well-suited to process sequential data as they capture
long-term dependencies. A seq2seq architecture is composed of two
components: an encoder and a decoder \cite{NIPS2014_5a18e133, cho-etal-2014-learning}. The encoder maps the input sequences into a latent representation and a context vector, which are passed to the decoder.

\textbf{Encoder\((E)\)}: The encoder processes fixed-length input
sequences of shape \((T,\ X)\). It consists of two unidirectional LSTM
layers; the first is followed by batch normalization. The number of
units in each layer was selected via hyperparameter tuning during model
training. The second LSTM layer generates two outputs: a latent
representation \(\left( H_{enc}^{(2)} \right)\), summarizing hidden
states across all time steps, and a context vector
\(\left( H_{enc_{f}}^{(2)} \right)\), representing the final hidden and
cell state.

\begin{equation}
\left( h_t^{(1)}, \ c_t^{(1)} \right) = LSTM_{ELU}^{(1)}\left( x_t, \ h_{t-1}^{(1)}, \ c_{t-1}^{(1)} \right), \quad t = 1, \ldots, T
\label{eq:lstm1}
\end{equation}

\begin{equation}
\left( h_t^{(2)}, \ c_t^{(2)} \right) = LSTM_{ELU}^{(2)}\left( \widetilde{h_t}^{(1)}, \ h_{t-1}^{(2)}, \ c_{t-1}^{(2)} \right), \quad t = 1, \ldots, T
\label{eq:lstm2}
\end{equation}

\begin{equation}
H_{enc}^{(2)} = \left( h_1^{(2)}, \ h_2^{(2)}, \ \ldots, \ h_T^{(2)} \right) \in \mathbb{R}^{T \times d}
\label{eq:henc}
\end{equation}

\begin{equation}
H_{enc_f}^{(2)} = \left( h_T^{(2)}, \ c_T^{(2)} \right) \in \mathbb{R}^{T \times d}
\label{eq:henc-final}
\end{equation}

\textbf{Decoder\((D)\)}: The decoder receives same input sequences
as the encoder, of shape \((T,\ X)\); consists of two unidirectional
LSTM layers. The initial hidden state of first layer is initialized with
\(H_{enc_{f}}^{(2)}\). 
\begin{figure}[ht]
    \centering
    \includegraphics[width=\textwidth]{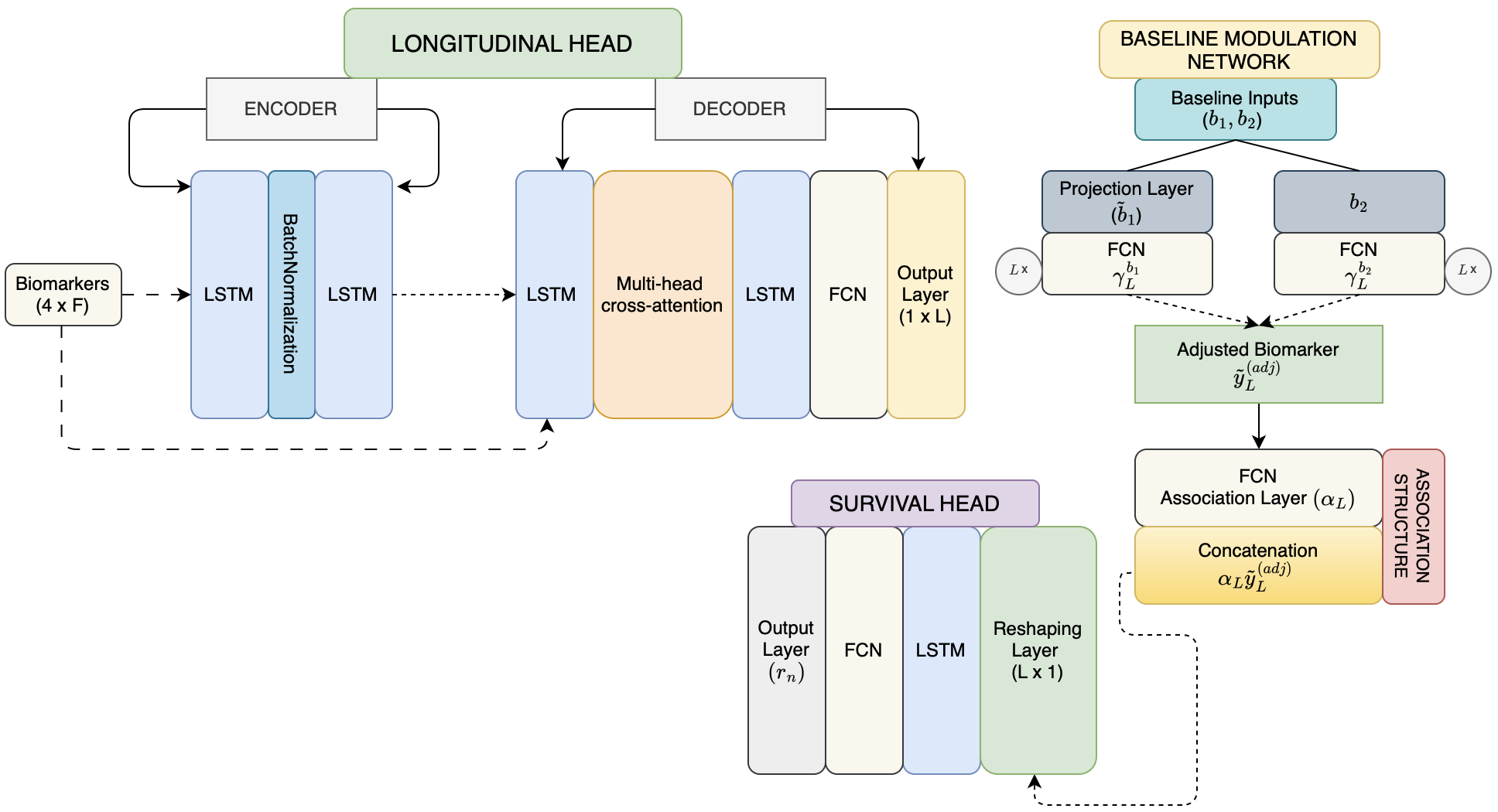}
    \caption{DeepAJM Model Architecture}
    \label{fig:deepajm-architecture}
\end{figure}
The output from this layer \(\left( H_{dec}^{(1)} \right)\) serves as the query \((Q)\), one of the three inputs requires by a multiheaded cross-attention layer. The attention layer is followed by a unidirectional LSTM layer and two dense layers -- the first with an ELU (exponential linear unit) activation function and the second with a linear activation function.

\begin{equation}
\left( h_t^{(dec,\,1)}, \ c_t^{(dec,\,1)} \right) = LSTM_{ELU}^{(dec,\,1)}\left( x_t, \ h_{t-1}^{(dec,\,1)}, \ c_{t-1}^{(dec,\,1)} \right), \quad t = 1, \ldots, T
\label{eq:decoder-lstm1}
\end{equation}

with
\[
h_0^{(dec,\,1)} = h_T^{(2)}, \quad c_0^{(dec,\,1)} = c_T^{(2)}
\]

\begin{equation}
H_{dec}^{(1)} = \left( h_1^{(dec,\,1)}, \ h_2^{(dec,\,1)}, \ \ldots, \ h_T^{(dec,\,1)} \right) \in \mathbb{R}^{T \times d}
\label{eq:hdec1}
\end{equation}

\subsubsection*{\textbf{Multi-headed cross-attention}}

Multi-head cross-attention establishes relationship between two
sequences by enabling one sequence to attend to key elements of the
other \cite{vaswani2017attention}. This attention mechanism enforces the model to extract salient information from \(H_{enc}^{(2)}\) to enrich the decoder's output. The layer received three inputs \(-\) query \((Q)\), key \((K)\), and value \((V)\); the query was derived from \(H_{dec}^{(1)}\), while
\(H_{enc}^{(2)}\ \) served as the key and value, as defined in Eq.\((8)\). The resulting attention scores from this layer was concatenated
with \(H_{dec}^{(1)}\) and passed to the second LSTM layer. The number
of heads was set to 2 and per-head dimension \(d_{k}\) to 16, selected
empirically after multiple trials. The multi-head attention scores and
the concatenation layer are given as:

\begin{equation}
Q_i = H_{dec}^{(1)} W_Q^{(i)}, \quad K_i = H_{enc}^{(2)} W_K^{(i)}, \quad V_i = H_{enc}^{(2)} W_V^{(i)}
\label{eq:qkv}
\end{equation}

\begin{equation}
head_i = Attention\left( Q_i, \ K_i, \ V_i \right), \quad i \in \{1, 2\}
\label{eq:attn-head}
\end{equation}

\begin{equation}
Z_{ct} = Concat\left( Multihead(Q, K, V), \ H_{dec}^{(1)} \right)
\label{eq:concat}
\end{equation}

\subsubsection{\textbf{Baseline-covariate modulation network}}

The baseline modulation network was incorporated to readjust the
longitudinal outputs by baseline covariates, such that each adjusted
output carries individual baseline-specific effects together with the
temporal patterns. Inputs to this network are processed baseline
covariates and the outputs generated by the longitudinal head. For each
\(L\) longitudinal output, baseline covariates are each passed through a
separate fully connected layer without activation, i.e., \(L\) parallel
projections per covariate, producing output-specific effects
\(\left( \gamma_{j}^{b_{1}},\ {\ \gamma}_{j}^{b_{2}} \right)\). The
adjusted longitudinal outputs are obtained as:

\begin{equation}
\widetilde{y}_{n,j(adj)} = y_{n,j} + \left( \gamma_j^{b_1} \cdot \widetilde{b}_{1,n} + c_j^{b_1} \right) + \left( \gamma_j^{b_2} \cdot b_{2,n} + c_j^{b_2} \right), \quad j = 1, \ldots, L; \ n = 1, \ldots, N
\label{eq:y-adjusted}
\end{equation}

\subsubsection{\textbf{Longitudinal-survival association structure}}

DeepAJM was devised with an interpretable, explicit association
structure that linked the longitudinal process to the survival process.
It assigns a unique associative weight to each adjusted longitudinal
output from the modulation network, generating an association-weighted
value per output. Each weight is output-specific and shared across
subjects. These association-weighted longitudinal outputs are
concatenated and passed to the event-process head, mapping to survival
risk scores.

\begin{equation}
s_{n,j} = \alpha_j \widetilde{y}_{n,j(adj)} + \beta_j, \quad j = 1, \ldots, L; \ n = 1, \ldots, N
\label{eq:scaled-shift}
\end{equation}

\begin{equation}
s_n = concat\left( s_{n,1}, \ s_{n,2}, \ldots, \ s_{n,L} \right)
\label{eq:concat-s}
\end{equation}

\subsubsection{\textbf{Survival head}}

The association-weighted outputs were reshaped into a sequence
\(s_{n}' \in \mathbb{R}^{L \times 1}\) and processed by an LSTM layer,
enabling the model to capture cross-biomarker interactions. The LSTM
layer was followed by two fully connected layers, with ReLU and linear
activations respectively, producing log-hazard score \(r^{te}\) at
\(t_{n,T}\). Survival probabilities were obtained at \(t_{pred}\) using
\(r^{te}\) and baseline hazard, as described below.

\begin{equation}
r_n = f_{surv}\left( s_n' \right)
\label{eq:rn-surv}
\end{equation}

\subsubsection*{\textbf{Baseline Cumulative Hazard}} As the survival component is cox-based and generates survival probabilities at \(t_{pred}\), the individual
relative risk scores required a shared baseline hazard, anchored to an
absolute time scale. Therefore, the Breslow estimator \cite{breslow1974covariance}, a non-parametric estimate of the survival curve, was considered to estimate the baseline cumulative hazard from training log-hazard scores, providing the time-anchored scale that \(r^{te}\) alone cannot provide. The baseline cumulative hazard was then converted to the baseline survival curve, shared across all subjects. The baseline survival curve, along with the predicted log-hazard scores, was used to compute survival probabilities
at the prediction horizon. The incorporation of the Breslow estimator is
what makes the proposed joint model a Cox-based neural joint model.

The Breslow estimator that yielded the survival probabilities,

\begin{equation}
\widehat{H}_0(t) = \sum_{t_i \leq t} \frac{d_i}{\sum_{j \in R_{tr}(t_j)} e^{r_j^{tr}}}
\label{eq:baseline-hazard}
\end{equation}

where $d_i$ = number of events at time $t_i$; $r_j^{tr}$ = predicted log-hazard score for subject $j$ from the training fold; $R_{tr}(t_i) = \{j : t_j \geq t_i\}$ is the risk set built from training subjects.

The baseline survival curve at time $t$:

\begin{equation}
\widehat{S}_0(t) = \exp\left( -\widehat{H}_0(t) \right)
\label{eq:baseline-survival}
\end{equation}

The survival probability for subject $j$ at time $t$:

\begin{equation}
\widehat{S}(t \mid r_j^{te}) = \widehat{S}_0(t)^{\exp\left( r_j^{te} \right)}
\label{eq:survival-prob}
\end{equation}

where $r_j^{te}$ = predicted log-hazard score for subject $j$ from the test fold.

\subsubsection*{\textbf{Survival outcomes at prediction horizon}} The survival function is a right-continuous step function \cite{clark2003survival}, a piecewise constant between consecutive event times. The survival probability at \(t_{pred}\) was therefore, obtained by evaluating the step function at that horizon, i.e., carrying forward the estimate from the last event time at or before that horizon. The earliest time point \(\left( t' \right)\) in the prediction horizon was set to year 5.

For the ordered distinct event times $t_1 < t_2 < t_3 < \ldots < t_j < \ldots < t_m$:

\begin{equation}
\widehat{S}(t') =
\begin{cases}
1, & t' < t_1 \\
\widehat{S}(t_j), & t_j \leq t' < t_{j+1}
\end{cases}
\label{eq:step-survival}
\end{equation}

\section{Experimental Setup}
\label{sec:experimental-setup}

\subsection{Datasets}
\label{sec:datasets}

\textbf{Cardiovascular Diseases (CVD):} This dataset contained electronic health
records (EHRs), obtained from the University of Kanas Health System with
Institutional Review Board (IRB) approval from The University of Kansas
Medical Center. The cohort included subjects with various cardiovascular
diseases recorded between 1980 and 2024, comprising longitudinal,
time-to-event and baseline covariate information. After preprocessing,
2257 subjects were retained from the cohort, each with a minimum of 5
and maximum of 20 clinical encounters. The longitudinal variables
included from this dataset were physiological and clinical biomarkers --
diastolic blood pressure, systolic blood pressure, basal metabolic index
(BMI) and respiration rate. The time at which all these biomarkers were
collectively sampled at each clinical encounter was also extracted and
used in the analyses. The time-to-event or time-to-censoring information
for each subject was computed based on the provided event time and event
indicator information.

\textbf{Primary biliary cirrhosis (PBC2):} This dataset contained longitudinal and time-to-event data along with baseline covariates for 312 subjects diagnosed with primary biliary cirrhosis \cite{murtaugh1994pbc}. The subjects were enrolled in a randomized controlled trial conducted by the Mayo Clinic between 1974 and 1984, and the dataset is publicly available. The dataset retained 183 subjects, each with a minimum of 5 and a maximum of 16 clinical encounters. The clinical and functional biomarkers used as longitudinal information were serum bilirubin, albumin, SGOT and prothrombin. The observation time for each clinical encounter was also included.

\textbf{Simulated dataset:} The data were simulated for 800 subjects with four
longitudinal variables and baseline covariates. The number of follow-up
visits per subject was sampled uniformly from \(\lbrack 4,10\rbrack\) ,
with visit times (in years) sampled irregularly from a continuous
uniform distribution on \(\lbrack 0,12\rbrack\), producing between 5 and
11 irregularly spaced encounters per subject. The longitudinal variables
were simulated using a nonlinear mixed-effects design. The
subject-specific random effects (intercept, slope and curvature) for all
variables jointly followed a multivariate normal distribution whose
variance-covariance structure induced correlation both within each
variable's subject-specific random effects and across variables. These
generated random-effects, combined with variable-specific fixed-effects
drawn from uniform distributions and gaussian measurement error
simulated the observed longitudinal measurements
\(\lbrack Eq.\ 22 \rbrack\). Baseline covariates were simulated once per
subject, with age following a gaussian distribution and sex a Bernoulli
distribution. A shared-random-effects joint model then linked the
underlying true (noise-free) latent trajectories to a Weibull baseline
hazard \((shape = 1.2)\), through a variable-specific association
parameter \(\alpha_{k}\), shared across subjects. Event times were
simulated using simsurv package in R \cite{simsurv}. After preprocessing, the
dataset retained 538 subjects, with a minimum of 5 and a maximum of 11
encounters per subject. For subject \(j\), the underlying trajectory for
\(k^{th}\) biomarker at time \(t\),

\begin{equation}
Y_{jk}(t) = \beta_{0k} + b_{0jk} + \left( \beta_{1k} + b_{1jk} \right)t + \left( \beta_{2k} + b_{2jk} \right)t^2 + \epsilon_{jk}, \quad k = 1,2,\ldots,4
\label{eq:longitudinal-submodel}
\end{equation}

\begin{equation}
Y_{jk}(t) = \mu_{jk}(t) + \epsilon_{jk}
\label{eq:y-mu-eps}
\end{equation}

\begin{equation}
b_{jk} = \left\{ b_{0jk}, \ b_{1jk}, \ b_{2jk} \right\}, \quad b_j \sim MVN(0, \Sigma)
\label{eq:random-effects}
\end{equation}

For subject $j$, the hazard function is:

\begin{equation}
h_j(t) = h_0(t) \exp\left( \sum_{k=1}^{4} \alpha_k \mu_{jk}(t) \right)
\label{eq:hazard}
\end{equation}

\begin{equation}
\alpha_k \sim Uniform(0.01, 0.03)
\label{eq:alpha-prior}
\end{equation}

\subsection{Model Training and Optimization}
\label{sec:training-optimization}
A 5-fold nested cross-validation with a 3-fold inner cross-validation
was employed to train and evaluate all models, including baseline
methods. Within each inner loop, the hyperparameters were tuned using
Bayesian optimization \cite{omalley2019kerastuner}, which selected the best configuration of hyperparameters based on the results of past trials. As a result, this optimization method required fewer trials than an exhaustive search to converge to the optimal set of hyperparameters. The number of
optimization trials varied by dataset and was specified in the
supplementary section. Prior to training, longitudinal features
including age were scaled to \(\lbrack 0,1\rbrack\) using scikit-learn's
MinMaxScaler \cite{pedregosa2011scikit}.

\textbf{Loss Functions}: The model was trained by jointly optimizing longitudinal
and survival loss. The longitudinal loss is the mean absolute error
between the observed and predicted longitudinal outputs, and cox
proportional-hazards partial-likelihood loss is defined as the survival
loss. The relative contribution of survival loss to joint objective is
controlled by a weighting coefficient \(\lambda\), which was tuned as an
additional hyperparameter.

\paragraph{Longitudinal Loss:}
\label{sec:longitudinal-loss}

\begin{equation}
L_{long} = \frac{1}{NL} \sum_{i=1}^{N} \sum_{k=1}^{L} \left| y_{i,k} - \widehat{y}_{i,k} \right|
\label{eq:l-long}
\end{equation}

\paragraph{Survival Loss:}
For each sample $i$, at time $t_i$ with $\delta_i \in \{0, 1\}$ as the event indicator, and the predicted log-hazard score $r_i^{te}$, the survival loss was computed as:

\begin{equation}
L_{surv} = -\frac{1}{N} \sum_{i=1}^{N} \delta_i \left[ r_i^{te} - \log \sum_{j \in R(t_i)} e^{r_j^{te}} \right]
\label{eq:l-surv}
\end{equation}

where $R(t_i) = \{j : t_j \geq t_i\}$ is the risk set at time $t_i$.

\paragraph{Joint Loss:}
The final training objective is the combination of both losses:

\begin{equation}
L_j = L_{long} + \lambda_{surv} \cdot L_{surv}
\label{eq:l-joint}
\end{equation}

\subsubsection{Evaluation metrics}\label{evaluation-metrics}

Four metrics were used to evaluate the performance of survival predictions: concordance index (C-index), integrated Brier score (IBS), time-dependent AUC (t-AUROC) and time-dependent AUPRC (t-AUPRC). The IBS and t-AUROC employed inverse probability of censoring weighting (IPCW) \cite{robins1992recovery}, as implemented in scikit-survival \cite{sksurv}, to account for right-censoring, with censoring weights estimated by the Kaplan-Meier (KM) method. The same KM-based IPCW was applied to account for censoring when computing t-AUPRC.

\subsubsection{\textbf{Calibration and metric computation}}

The Breslow estimator produces survival probabilities by imposing the
proportional hazards assumption, and therefore, maintains a relative
risk ranking shared across all prediction horizons. However, IBS is
sensitive to absolute predicted values, so calibration of survival
probabilities is essential for computing this metric. To address that,
the flexible hazard regression model (hare) \cite{polspline} from the polspline package in R \cite{kooperberg1995hazard} was used to calibrate the predicted survival probabilities at the prediction horizon. As hare models the hazard as a flexible function of time and risk scores, it relaxes the proportional hazards assumption and causes the ranking order to vary across horizons. As a result,
uncalibrated survival probabilities by Breslow's were used to compute
discrimination-sensitive metrics, C-index and t-AUROC. However,
calibrated survival probabilities by hare improved t-AUPRC, which showed
its greater sensitivity to ranking at the high-precision region.
Therefore, calibrated probabilities were used for IBS and t-AUPRC, and
uncalibrated survival probabilities for C-index and t-AUROC.

\subsection{Baseline Methods}
\label{sec:baseline-methods}

Our proposed model was compared against four baselines: the parametric
joint model, the TransformerJM model, the DA-LSTM model and Cox-based
survival-only model. All models were trained and evaluated under
identical conditions: same inputs, data splits, outer folds, inner
cross-validation for hyperparameter optimization, and evaluation
metrics.

\textbf{Parametric Joint Model (PJM):} This baseline comprised of two submodels:
a Cox proportional hazard model as the survival submodel and a linear
mixed-effects as the longitudinal submodel \cite{rizopoulos2012joint}. The model was trained and
evaluated using the JMBayes2 package \cite{jmbayes2} in R.

\textbf{TransformerJM}: TransformerJM jointly models longitudinal and time-to-event data using a sequence-to-sequence architecture \cite{TFJM}. The author's implementation was used and was evaluated under identical conditions as our proposed model.

\textbf{DA-LSTM:} This method was proposed as a deep learning approach for jointly modeling longitudinal and time-to-event data \cite{pmlr-v85-lim18a}. The used both a recurrent neural network (RNN) with ELU activation function
(DA-NN) and a long short-term memory (LSTM) network (DA-LSTM) as the
shared temporal layer within the same architecture. Upon evaluation,
they selected DA-LSTM as the primary model because it outperformed
DA-NN. Accordingly, DA-LSTM was adopted as the baseline from comparison
with our proposed model. Since neither DA-LSTM nor DA-NN architecture
incorporated time of longitudinal measurements as an additional input,
and to evaluate DA-LSTM under identical conditions as our proposed
model, we added time as additional feature alongside longitudinal and
baseline features, yielding DA-LSTM(t), a reimplementation that
incorporates time as the time elapsed from subject's baseline visit. The
author's original implementation was not publicly available, and
therefore, we reproduced it following the descriptions in their paper.
The implementation details for DA-LSTM are provided in the supplementary
section.

\textbf{Cox-based survival-only (Surv-only):} This baseline model consisted of
the survival head, the same survival head as in the proposed model but
without the LSTM layer as it received only baseline features with no
longitudinal features as inputs. The baseline features were processed
and trained in the same manner as done in our proposed joint model, but
the model was trained using only the survival loss.

\section{Results}\label{results}

DeepAJM was compared against four methods: a parametric joint model
(PJM), a transformer-based joint model (TransformerJM), an RNN-based
joint model (DA-LSTM) and a Cox-based survival-only model. All methods
were evaluated under an identical protocol across two real-world
datasets and one simulated dataset. For each method, survival
predictions were obtained at the \(t\_ pred\) horizon, with the first
evaluated time point at \(t' = 5\), and subsequent evaluation at
\(t' + 1,\ t' + 2\). \\

\begin{table}[ht]
\centering
\caption{Concordance index (C-index) across datasets for DeepAJM and benchmarked models.}
\label{tab:cindex}
\resizebox{\textwidth}{!}{%
\begin{tabular}{@{}lcccccc@{}}
\toprule
Dataset & DeepAJM & PJM & TransformerJM & DA-LSTM & DA-LSTM (t) & Surv-only \\
\midrule
CVD       & \textbf{0.9114} & 0.5124 & 0.6609 & 0.6621 & 0.8474 & 0.6166 \\
PBC2      & \textbf{0.8073} & 0.5031 & 0.5876 & 0.7188 & 0.6817 & 0.5543 \\
Simulated & 0.6410          & \textbf{0.6422} & 0.5897 & 0.5895 & 0.6316 & 0.4609 \\
\bottomrule
\end{tabular}%
}
\end{table}

\begin{table}[ht]
\centering
\caption{Integrated Brier Score (IBS) across datasets for DeepAJM and benchmarked models.}
\label{tab:ibs}
\resizebox{\textwidth}{!}{%
\begin{tabular}{@{}lcccccc@{}}
\toprule
Dataset & DeepAJM & PJM & TransformerJM & DA-LSTM & DA-LSTM (t) & Surv-only \\
\midrule
CVD       & \textbf{0.1122} & 0.3070 & 0.2592 & 0.2138 & 0.1733 & 0.2203 \\
PBC2      & \textbf{0.1012} & 0.1584 & 0.1322 & 0.5895 & 0.3860 & 0.1360 \\
Simulated & 0.1330          & \textbf{0.0743} & 0.1107 & 0.1342 & 0.1956 & 0.1066 \\
\bottomrule
\end{tabular}%
}
\end{table}

Each reported metric is the average across all evaluated time points from the 5-fold nested cross-validation. On both real-world datasets, DeepAJM achieved the highest C-index, t-AUROC, t-AUPRC, and the lowest IBS. On the simulated dataset, it outperformed all baselines on t-AUROC and t-AUPRC, and achieved a C-index equivalent to best-perfomring baseline (PJM). Further, DA-LSTM (t) showed improved C-index, t-AUROC and t-AUPRC over the original DA-LSTM on the CVD and simulated datasets.

\begin{table}[ht]
\centering
\caption{Time-dependent AUROC across datasets for DeepAJM and benchmarked models.}
\label{tab:auroc}
\resizebox{\textwidth}{!}{%
\begin{tabular}{@{}lcccccc@{}}
\toprule
Dataset & DeepAJM & PJM & TransformerJM & DA-LSTM & DA-LSTM (t) & Surv-only \\
\midrule
CVD       & \textbf{0.9374} & 0.5127 & 0.6884 & 0.7028 & 0.8898 & 0.6190 \\
PBC2      & \textbf{0.8878} & 0.4497 & 0.6600 & 0.7758 & 0.7928 & 0.5623 \\
Simulated & \textbf{0.7483} & 0.5551 & 0.5034 & 0.5194 & 0.6876 & 0.4333 \\
\bottomrule
\end{tabular}%
}
\end{table}

\begin{table}[H]
\centering
\caption{Time-dependent AUPRC across datasets for DeepAJM and benchmarked models.}
\label{tab:auprc}
\resizebox{\textwidth}{!}{%
\begin{tabular}{@{}lcccccc@{}}
\toprule
Dataset & DeepAJM & PJM & TransformerJM & DA-LSTM & DA-LSTM (t) & Surv-only \\
\midrule
CVD       & \textbf{0.8548} & 0.3635 & 0.5500 & 0.5777 & 0.7988 & 0.4479 \\
PBC2      & \textbf{0.5270} & 0.1752 & 0.3206 & 0.4090 & 0.4490 & 0.1687 \\
Simulated & \textbf{0.2462} & 0.1046 & 0.1276 & 0.1527 & 0.2201 & 0.1261 \\
\bottomrule
\end{tabular}%
}
\end{table}

\section{Discussion}

DeepAJM is a deep joint model that decodes complex temporal dynamics of
a patient's health profile with an LSTM-based seq2seq architecture. A
baseline-covariate modulation network captures the distinct, implicit
effects of individual baseline covariates such as age and sex,
readjusting each longitudinal outcome by covariate-specific weights, so
that it carries richer information for downstream analysis. An explicit
longitudinal-survival association structure then generates association
weights, assigns them to the corresponding adjusted longitudinal
outcome, and maps to a Cox-based survival head, which generates
log-hazard scores that are combined with the baseline hazard curve to
obtain survival probabilities at prediction horizon. DeepAJM achieved
the strongest overall performance across all datasets and metrics,
outperforming the baselines on both discrimination and calibration. It
attained a consistently higher C-index on two real-world datasets and
matched the best baseline on simulated dataset, showing DeepAJM's robust
preservation of survival-risk ranking. For calibration, lower IBS values
showed improved calibration and consistently higher t-AUROC and t-AUPRC
values indicated stronger discrimination for both event and event-free
subjects. Its higher t-AUPRC at horizons with low event rates reflects
DeepAJM's accuracy in the regime where survival models struggle most.

The purpose of jointly modeling longitudinal and survival processes is
to enhance the predictive power of survival outcomes by accounting for
hidden, unobserved factors that potentially influence the risk scores.
Built on a LSTM-based sequence-to-sequence network, the model strongly
captured temporal patterns present in heterogenous longitudinal
trajectories across the population. Integrating these temporal latent
representations with the baseline covariates further enriched the
longitudinal outcomes, since baseline covariates tend to affect
subject's longitudinal trajectories differently. To realize the gains in
preserving the interpretable structure while simultaneously enhancing
model's predictive power, the survival head was implemented as a neural
network-based cox model and trained on cox partial likelihood loss.

The latent trajectory of longitudinal outcomes was linked to the
survival process through an association structure, representing the
proposed model a semi-parametric joint model. As the survival submodel
further employed an LSTM, it learned a cross-biomarker structure from
the association-weighted outcomes. To obtain survival probabilities at
fixed prediction horizons, the baseline cumulative hazard was computed
from training subjects using Breslow estimator \cite{breslow1974covariance}. Since the Breslow estimator relies on proportional hazards (PH) assumption, we verified this assumption, and the results indicated no violation of PH assumption
on both input covariates and log-hazard scores. Consequently, the
proposed model shared structural and partial interpretability properties
of the classical parametric joint model while also addressing its
constraints. Overall, the model achieved robust discrimination and
calibration as compared to the classical parametric joint model while
overcoming several of its limitations.

Prior work on joint modeling of longitudinal and survival outcomes \cite{TFJM, zhang2025transformerlsr, pmlr-v85-lim18a} did not incorporate an association structure, a key mechanism by which a joint model learns how the longitudinal process informs the survival process. TransformerJM \cite{TFJM} employed a transformer-based architecture to predict survival probabilities at prediction horizons and then autoregressively fed these predictions back to predict outcomes at subsequent time points. Their architecture jointly modeled
longitudinal and survival data together with baseline covariates and
time but it did not learn any explicit association structure linking the
longitudinal outcomes to the survival process. Additionally, it
discretized event times into fixed bins, which can yield highly skewed
survival estimates in bins with fewer events. Thus, although this
transformer-based baseline jointly modeled the longitudinal and survival
outcomes, it did not explicitly maintain the inherent properties of a
joint model. DA-LSTM \cite{pmlr-v85-lim18a} used an RNN-based architecture to process temporal information and model the hazard rate through exponential distribution, producing piecewise-constant hazard. It is multi-output model; since our study focused solely on survival predictions, we obtained only the
survival output and considered it as a baseline. Like TransformerJM,
DA-LSTM did not learn any association structure between the longitudinal
and survival processes, which is an explicit gap that our proposed model
addressed. Furthermore, the original DA-LSTM did not handle the
irregular gap between sampled observations, which is one of the
challenges in EHR datasets. To evaluate it as baseline under conditions
identical to our model, DA-LSTM was additionally trained with time of
longitudinal measurements as an input. As their original implementation
of DA-LSTM did not include time, we reported DA-LSTM's performance both
with and without measurement time, and the results showed notable
difference between two settings. This evidently showed that in
real-world datasets with irregular sampling intervals, incorporating
measurement times of longitudinal observations enabled the model to
capture temporal distance between irregularly sampled observations and
enhanced its predictive ability.

The model produced cox-style nonlinear log-hazard scores that were
exponentiated to compute the baseline hazard curve, and therefore, the
ordering of hazards across subjects were preserved over time, which can
further be stratified into risk groups. Therefore, the model attained
discriminative interpretability as well as an estimable population
baseline curve whose shape over time can be studied for clinical
interpretation. The model also incorporated a per-biomarker association
layer, analogous to classical joint model's association structure, and
which generated outcome-specific association weights that are
extractable and interpretable. However, as these adjusted longitudinal
outcomes passed through multiple nonlinear layers downstream, they did
not correspond to closed-form hazard ratios as a in parametric joint
model.

The architecture was designed to predict the log-hazard at
\(T_{n,\ \tau}\); however, as we evaluated the model performance at
fixed prediction horizons for all subjects, the baseline cumulative
hazard along with the predicted risk scores were used to compute
survival probabilities on the event times grid. The survival function, a
right-continuous step function, was used to obtain the outcomes at those
horizons. This restricts the model to predicting survival outcomes at
horizons that fall at or before the last observed event time. Although
the model can in principle predict outcomes at any horizon given the
history up to the landmarking time, the requirement to evaluate at fixed
horizons for irregularly sampled clinical observations made it necessary
to employ a shared baseline hazard, a limitation that also required
verifying the PH assumption. Our model used four time points as
longitudinal inputs, but it can accommodate any number of time points.

The joint model carries greater clinical relevance for disease
progression prediction than survival-only models, as it explores the
underlying biological process that often reflects latent factors driving
event occurrence. It thereby enables producing dynamic, individualized
predictions and that can be updated at the point of care by forecasting
the longitudinal trajectory.

\clearpage
\bibliographystyle{plainnat}   
\bibliography{references}     
\end{document}